\documentclass[twocolumn,
superscriptaddress,
showpacs,preprintnumbers,amsmath,amssymb]{revtex4}
\usepackage{graphicx}% Include figure files

\usepackage{amssymb,amsmath,amsthm}
\usepackage{epsfig}
\newtheorem{Thm}{Theorem}
\newtheorem{Lem}{Lemma}
\newtheorem{Prop}{Proposition}

\newcommand{\ket}[1]{{\left| #1 \right\rangle}}

\newcommand{\Z}{\mbox{$\mathbb Z$}}

\newcommand{\cI}{\mathcal{I}}

\newcommand{\cF}{\mathcal{F}}

\begin{document}
%%%%%%%%%%%%%%%%%%%%%%%%%%%%%%%%%%%%%%%%%%%%%%%%%%%%%%%%%%%%%%%%%%%%%%%%%%
%                                                                        %
%                                 Title                                  %
%                                                                        %
%%%%%%%%%%%%%%%%%%%%%%%%%%%%%%%%%%%%%%%%%%%%%%%%%%%%%%%%%%%%%%%%%%%%%%%%%%
\title{Notes on the hidden subgroup problem on some semi-direct product groups}

\author{Dong Pyo Chi}\email{dpchi@math.snu.ac.kr}
\affiliation{
 Department of Mathematical Sciences and Research Institute of Mathematics,
 Seoul National University, Seoul 151-742, Korea
}
\author{Jeong San Kim}\email{freddie1@snu.ac.kr}
\affiliation{
 Department of Mathematical Sciences and Research Institute of Mathematics,
 Seoul National University, Seoul 151-742, Korea
}
\author{Soojoon Lee}\email{level@khu.ac.kr}
\affiliation{
 Department of Mathematics and Research Institute for Basic Sciences,
 Kyung Hee University, Seoul 130-701, Korea
}
\date{\today}

%%%%%%%%%%%%%%%%%%%%%%%%%%%%%%%%%%%%%%%%%%%%%%%%%%%%%%%%%%%%%%%%%%%%%%%%%%
%                                                                        %
%                              Abstract                                  %
%                                                                        %
%%%%%%%%%%%%%%%%%%%%%%%%%%%%%%%%%%%%%%%%%%%%%%%%%%%%%%%%%%%%%%%%%%%%%%%%%%
\begin{abstract}
We consider the hidden subgroup problem
on the semi-direct product of cyclic groups $\Z_{N}\rtimes\Z_{p}$
with some restriction on $N$ and $p$.
By using the homomorphic properties,
we present a class of semi-direct product groups
in which the structures of subgroups can be easily classified.
Furthermore, we show that
there exists an efficient quantum algorithm
for the hidden subgroup problem on the class.
\end{abstract}

\pacs{
03.67.Lx, % Quantum computation
02.20.Bb  % General structures of groups
%03.65.Ta % Foundations of quantum mechanics; measurement theory
}
%\keywords{}
\maketitle

%%%%%%%%%%%%%%%%%%%%%%%%%%%%%%%%%%%%%%%%%%%%%%%%%%%%%%%%%%%%%%%%%%%%%%
%%%                                                                %%%
%%%                         Introduction                           %%%
%%%                                                                %%%
%%%%%%%%%%%%%%%%%%%%%%%%%%%%%%%%%%%%%%%%%%%%%%%%%%%%%%%%%%%%%%%%%%%%%%

\section{Introduction}
Most of exponentially fast quantum algorithms,
such as the Simon algorithm~\cite{Simon} and the period-finding algorithm~\cite{Shor},
can be regarded as one for a special problem called the hidden subgroup problem (HSP).
%which is to find a subgroup hidden by an oracle function defined on a given group.
The HSP is to find a subgroup $H$ of a given group $G$ with an oracle function $f$ defined on $G$
such that $f(a)=f(b)$ if and only if $aH=bH$ for all $a$, $b$ in $G$.
It is well known that
if the group $G$ is abelian then the HSP on $G$ can be efficiently solved
by a quantum algorithm of running time polynomial in $\log|G|$,
while no solution is known for the general case of nonabelian groups.
In particular,
since the graph isomorphism problem and certain lattice problem can be reduced
to the HSP on the symmetric group and the HSP on the dihedral group, respectively~\cite{iso,lattice},
it was naturally asked whether there exists
an efficient quantum algorithm for the HSP on nonabelian groups,
and has been actively
studied~\cite{ettinger,HRT,GSVV,friedl,kuperberg,EHN,gavinsky,inui,moore,regev,BCD,BCD2}.

One way to construct a quantum algorithm for the HSP
is first to explicitly investigate the structures of all subgroups of a given group,
and then to find a quantum algorithm applicable to each subgroup structure.
Recently, Inui and Le Gall~\cite{inui} presented an efficient quantum algorithm for
the HSP on the groups $\Z_{p^r}\rtimes\Z_p$ for odd prime $p$
by classifying all the possible subgroups.
Employing such a method, we can also show that there exists an efficient quantum algorithm
for the HSP on $\Z_{2p^r}\rtimes\Z_p$ for odd prime $p$ as in Appendix.

Since multiplicative groups $\Z_{p^r}^*$ and $\Z_{2p^r}^*$ are cyclic,
it can be verified that $\Z_{p^r}\rtimes\Z_p$ and $\Z_{2p^r}\rtimes\Z_p$
have the same form of subgroup structures,
and hence we can obtain the same result for the HSP.
However, if one exploits the above method to solve the HSP on general semi-direct product groups,
then it is hard to find a quantum algorithm for the HSP on the groups
since it is difficult, even mathematically, to classify the subgroup structures.

In this work,
we consider the HSP on the semi-direct products of cyclic groups $\Z_{N}\rtimes\Z_{p}$,
where $N$ is factorized as $N=p_1^{r_1}p_2^{r_2}\cdots p_n^{r_n}$,
and a prime $p$ does not divide each $p_j-1$.
By using the homomorphic properties,
we show that there exists an efficient quantum algorithm for the HSP on the groups.

%%%
%%%    organization of this paper
%%%
This paper is organized as follows.
In Sec.~\ref{Sec:Semi-Direct}
we briefly introduce the semi-direct product groups, and
%In Sec.~\ref{Sec:Auto}
explain some homomorphic properties of the semi-direct product groups.
In Sec.~\ref{Sec:Algo2}
we show that there exists an efficient quantum algorithm for the HSP on the groups.
Finally, in Sec.~\ref{Sec:summary} we summarize our result.

%%%%%%%%%%%%%%%%%%%%%%%%%%%%%%%%%%%%%%%%%%%%%%%%%%%%%%%%%%%%%%%%%%%%%%%%%
%%%                                                                   %%%
%%%                semi-direct product                                %%%
%%%                                                                   %%%
%%%%%%%%%%%%%%%%%%%%%%%%%%%%%%%%%%%%%%%%%%%%%%%%%%%%%%%%%%%%%%%%%%%%%%%%%
\section{Semi-direct product groups}\label{Sec:Semi-Direct}
%\begin{Def}\label{Def:semi-direct product}
For any positive integer $N$ and $p$, and any group homomorphism $\phi$
from the group $\Z_{p}$ into the group of automorphisms of $\Z_{N}$,
the semi-direct product group $\Z_{N}\rtimes_{\phi}\Z_{p}$
is the set $\{(a,b): a \in \Z_{N}, b\in \Z_{p}\}$
with the group operation
$(a_{1},b_{1})(a_{2},b_{2})=(a_{1}+\phi(b_{1})(a_{2}),~b_{1}+b_{2})$.
%\end{Def}

We note that the group $\Z_N\rtimes_\phi\Z_p$
is obviously generated by the two elements $x=(1,0)$ and $y=(0,1)$,
and that $\phi$ is completely determined by the function value $\phi(1)(1)$,
since $\phi:\Z_p\rightarrow\mathrm{Aut}(\Z_N)$ is a homomorphism
and $\phi(a)$ is an automorphism for every $a\in\Z_p$.
Using the fact that
$\phi(b)(a)=a\phi(1)(1)^b$, we obtain the relation
\begin{equation}
y^{b}x^{a}=x^{a\phi(1)(1)^{b}}y^{b}.
\label{eq:ab_relation}
\end{equation}

Due to the facts that $\phi(1)(1)$ is relatively prime to $N$
and that $\phi(1)(1)^p= 1\pmod N$,
$\Z_N\rtimes_\phi\Z_p$ is completely determined by $\phi(1)(1)$.
For example, the case $\phi(1)(1)=1$ %is a trivial one which
leads to the direct product $\Z_N\times\Z_p$, and
if $\phi(1)(1)\neq 1$ and $p$ is prime then  $p$ is the smallest positive integer
satisfying
\begin{equation}
\phi(1)(1)^p= 1\pmod N,
\label{eq:order_p}
\end{equation}
that is,
$\phi(1)(1)$ is one of elements of $\Z_{N}^{*}$ with order $p$,
and $p$ is a divisor of $\varphi(N)$, where $\varphi$ is the Euler phi-function.
%where $\Z_{n}^{*}$ is the group consisting of the element in $\Z_{n}$
%which are relatively prime to $n$.

%%%%%%%%%%%%%%%%%%%%%%%%%%%%%%%%%%%%%%%%%%%%%%%%%%%%%%%%%%%%%%%%%%%%%%%%%
%%%                                                                   %%%
%%%                Notes on automorphisms                             %%%
%%%                                                                   %%%
%%%%%%%%%%%%%%%%%%%%%%%%%%%%%%%%%%%%%%%%%%%%%%%%%%%%%%%%%%%%%%%%%%%%%%%%%
%\section{Notes on automorphisms}\label{Sec:Auto}
We now consider the semi-direct product group $\Z_{q^{s}p^{r}}\rtimes_{\phi}\Z_{p}$,
where $p$ and $q$ are distinct primes, and $s$ and $r$ are positive integers.
%Then since $p$ is a divisor of $\varphi(q^{s}p^{r})=q^{s-1}p^{r-1}(q-1)(p-1)$,
Since $\Z_{q^{s}p^{r}}\rtimes_{\phi}\Z_{p}$ is isomorphic to
$(\Z_{q^{s}}\times \Z_{p^{r}})\rtimes_{\phi}\Z_{p}$,
for each $\alpha\in\Z_{p}$,
$\phi(\alpha)$ can be regarded as an automorphism on $\Z_{q^{s}} \times \Z_{p^{r}}$
such that $\phi(1)(1)$ is an element in $\Z_{q^{s}}^* \times \Z_{p^{r}}^{*}$ of order $p$.
Then we can obtain the following lemma:
\begin{Lem}\label{Lem:note1}
For each $\alpha\in\Z_{p}$, $a\in \Z_{q^{s}}$ and $b\in \Z_{p^{r}}$,
$\phi(\alpha)(a,0)=(a',0)$ and $\phi(\alpha)(0,b)=(0,b')$
for some $a'\in\Z_{q^{s}}$ and $b'\in \Z_{p^{r}}$.
\end{Lem}
\begin{proof}
Suppose $\phi(\alpha)(1,0)=(c,d)$ for some $c\in \Z_{q^{s}}$ and $d\in \Z_{p^{r}}$.
Then since $\phi(\alpha)$ is an homomorphism,
$\phi(\alpha)(0,0)=(0,0)$ and
\begin{eqnarray}
\phi(\alpha)(0,0)&=&\phi(\alpha)(q^{s},0)\nonumber\\
&=&q^{s}\phi(\alpha)(1,0)\nonumber\\
&=&(q^{s}c,q^{s}d).
\label{eq:phi_alpha00}
\end{eqnarray}
Since $\gcd(p,q)=1$ and $q^{s}d = 0 \pmod{p^{r}}$,
$d = 0 \pmod{p^{r}}$ and $\phi(\alpha)(1,0)=(c,0)$.
It follows that for any $a\in \Z_{q^{s}}$
$\phi(\alpha)(a,0)=(a',0)$ for some $a'\in\Z_{q^{s}}$.
Similarly, we can also obtain that for any $b\in \Z_{p^{r}}$,
$\phi(\alpha)(0,b)=(0,b')$ for some $b'\in \Z_{p^{r}}$.
\end{proof}

\begin{Lem}\label{Lem:note2}
Let $p$ and $q$ be distinct primes satisfying \mbox{$p\nmid q-1$}.
Then
\begin{equation}
(\Z_{q^{s}}\times\Z_{p^{r}})\rtimes_{\phi}\Z_{p}=
\Z_{q^{s}}\times (\Z_{p^{r}}\rtimes_{\psi}\Z_{p})
\label{eq:Lem2}
\end{equation}
for some homomorphism $\psi$ from $\Z_{p}$ to $\mathrm{Aut}(\Z_{p^{r}})$.
\end{Lem}
\begin{proof}
Since $\phi(p)=\phi(0)$ is the identity map $\cI$ on $\Z_{q^{s}}\times \Z_{p^{r}}$,
by Lemma~\ref{Lem:note1},
we can see
\begin{equation}
(1,0)=\cI(1,0)= \phi(p)(1,0)=\phi(1)^{p}(1,0)=(a^{p},0),
\label{eq:phi1p}
\end{equation}
where $(a,0)=\phi(1)(1,0)$ and $a^{p}= 1\pmod{q^{s}}$.
Since the order of $\Z_{q^{s}}^{*}$ is $q^{s-1}(q-1)$ and $p\nmid q-1$,
we obtain that $a$ must be $1$, that is, $\phi$ trivially acts on $\Z_{q^{s}}$.
Thus, for each $\alpha\in\Z_p$,
$\phi(\alpha)=\cI_0\times \psi(\alpha)$,
where $\cI_0$ is the identity map on $\Z_{q^{s}}$ and
$\psi$ is a homomorphism from $\Z_{p}$ to $\mathrm{Aut}(\Z_{p^{r}})$.

Therefore, the operation of the semi-direct product group is as follows:
\begin{eqnarray}
((a,b),c)((a',b'),c')&=&((a,b)+\phi(c)(a',b'),c+c')\nonumber\\
&=&(a+a',b+\psi(c)(b'),c+c'),\nonumber\\
\label{eq:operation_sdpg}
\end{eqnarray}
and this implies that Eq.~(\ref{eq:Lem2}) holds.
\end{proof}

 %\textbf{Note 3}~
\begin{Lem}\label{Lem:note3}
Let $G_0$ and $G_1$ be finite groups such that
the order of $G_0$ is relatively prime to the order of $G_1$,
and let $H$ be a subgroup of $G_0 \times G_1$.
Then
$H=H_0\times H_1$ for some subgroups $H_0$ of $G_0$ and $H_1$ of $G_1$.
\end{Lem}
\begin{proof}
Let $H$ be a subgroup of $G_0 \times G_1$ of order $rs$,
where $r$ and $s$ divide the order of $G_0$ and the order of $G_1$, respectively.
Then it is trivial that $r$ and $s$ are relatively prime to each other.
Now, for each $j=0, 1$, let $\pi_{j}$ be the natural projection from $G_0\times G_1$
onto $G_j$, and let $H_j=\pi_{j}(H)\subset G_j$.
Then we can readily show that $H = H_0\times H_1$.
\end{proof}

%%%%%%%%%%%%%%%%%%%%%%%%%%%%%%%%%%%%%%%%%%%%%%%%%%%%%%%%%%%%%%%%%%%%%%%%%
%%%                                                                   %%%
%%%                Quantum algorithms                                 %%%
%%%                                                                   %%%
%%%%%%%%%%%%%%%%%%%%%%%%%%%%%%%%%%%%%%%%%%%%%%%%%%%%%%%%%%%%%%%%%%%%%%%%%
\section{Quantum algorithm %for $\Z_{N}\rtimes_{\phi}\Z_{p}$
}\label{Sec:Algo2}
In this section, we present an efficient quantum algorithm
that can solve the HSP over the semi-direct product of cyclic groups
$\Z_{N}\rtimes_{\phi}\Z_{p}$,
which is not a direct product of $\Z_{N}$ and $\Z_{p}$,
where $N$ is factorized as $N=p_1^{r_1}p_2^{r_2}\cdots p_n^{r_n}$,
and a prime $p$ does not divide each $p_j-1$.

Since the $p_j$'s are all distinct primes and
$p$ is a divisor of $\varphi(N)=p_1^{r_1-1}p_2^{r_2-1}
\cdots p_n^{r_n-1}(p_1-1)(p_2-1)\cdots(p_n-1)$,
$p=p_k$ and $r_k\ge 2$ for some $k\in\{1, 2, \ldots, n\}$.
Thus, $N$ is a multiple of $p^2$.

Due to the factorization of $N$,
$\Z_{N}$ is isomorphic to the direct product of cyclic groups
$\Z_{p_{1}^{r_{1}}} \times \Z_{p_{2}^{r_{2}}} \times \cdots \times \Z_{p_{n}^{r_{n}}}$,
and we have
\begin{equation}
\Z_{N}\rtimes_{\phi}\Z_{p} \cong
(\Z_{p_{1}^{r_{1}}} \times \Z_{p_{2}^{r_{2}}} \times
\cdots  \times \Z_{p_{n}^{r_{n}}}) \rtimes_{\phi}\Z_{p}.
\label{eq:isomorphic}
\end{equation}
By Lemma~\ref{Lem:note1},
for each $\alpha\in\Z_p$,
the automorphism $\phi(\alpha)$ on
$\Z_{p_{1}^{r_{1}}} \times \Z_{p_{2}^{r_{2}}} \times \cdots  \times \Z_{p_{n}^{r_{n}}}$
acts trivially on each component of $\Z_{p_{j}^{r_{j}}}$ such that  $p$ differs $p_{j}$.
Let $\cI_1$ be the identity  on
$\Z_{p_{1}^{r_{1}}}  \times\cdots  \times \Z_{p_{k-1}^{r_{k-1}}}$
and $\cI_2$ be the identity on
$\Z_{p_{k+1}^{r_{k+1}}}  \times\cdots  \times \Z_{p_{n}^{r_{n}}}$.
Then, by Lemma~\ref{Lem:note2},
we obtain
\begin{equation}
\Z_{N}\rtimes_{\phi}\Z_{p}\cong
\Z_{p_{1}^{r_{1}}} \times \cdots \times
\widehat{\Z_{p_{k}^{r_{k}}}}\times \cdots  \times
 \Z_{p_{n}^{r_{n}}}\times (\Z_{p_{k}^{r_{k}}} \rtimes_{\psi}\Z_{p}),
 \label{eq:main}
\end{equation}
where $\Z_{p_{1}^{r_{1}}} \times \Z_{p_{2}^{r_{2}}} \times\cdots  \times
\widehat{\Z_{p_{k}^{r_{k}}}}\times \cdots  \times  \Z_{p_{n}^{r_{n}}} $
is the direct product of $\Z_{p_{j}^{r_{j}}}$'s except $\Z_{p_{k}^{r_{k}}}$
and $\psi$ is a homomorphism from $\Z_p$ to $\mathrm{Aut}(\Z_{p_{k}^{r_{k}}})$
such that $\phi(\alpha)=\cI_1\times \psi(\alpha)\times \cI_2$
for each $\alpha\in\Z_p$.
Moreover, since $\Z_{p_{1}^{r_{1}}} \times \Z_{p_{2}^{r_{2}}} \times\cdots  \times
\widehat{\Z_{p_{k}^{r_{k}}}}\times \cdots  \times  \Z_{p_{n}^{r_{n}}} $ is
a cyclic group of order $N/p_k^{r_k}$,
\begin{equation}
\Z_{N}\rtimes_{\phi}\Z_{p}\cong
\Z_{N/p_k^{r_k}}\times (\Z_{p_{k}^{r_{k}}} \rtimes_{\psi}\Z_{p}),
\end{equation}
by Eq.~(\ref{eq:main}).
Hence, the HSP on
$\Z_{N}\rtimes_{\phi}\Z_{p}$ is essentially equivalent to the HSP on
$\Z_{N/p_k^{r_k}}\times (\Z_{p_{k}^{r_{k}}} \rtimes_{\psi}\Z_{p})$.

Furthermore, since the order of $\Z_{N/p_k^{r_k}}$
is relatively prime to the order of the group $\Z_{p^{r_{k}}}\rtimes_{\psi}\Z_{p}$,
by Lemma~\ref{Lem:note3}, any subgroup $H$ of
$\Z_{N/p_k^{r_k}}\times (\Z_{p_{k}^{r_{k}}} \rtimes_{\psi}\Z_{p})$
is of the form $H_1 \times H_2$, where
$H_{1}$ and $H_2$ are subgroups of $\Z_{N/p_k^{r_k}}$
and $\Z_{p^{r_{k}}}\rtimes_{\psi}\Z_{p}$, respectively.
We note that the HSP on a cyclic group and the HSP on a group of the form
$\Z_{p^{r_{k}}}\rtimes_{\psi}\Z_{p}$ can be efficiently solved
by quantum algorithms~\cite{Shor,inui}.
Therefore, we can obtain the following result:
\begin{Thm}\label{Thm:main}
There exists an efficient quantum algorithm for the HSP on
$\Z_{N}\rtimes_{\phi}\Z_{p}$,
where $N$ is factorized as $N=p_1^{r_1}p_2^{r_2}\cdots p_n^{r_n}$,
and a prime $p$ does not divide each $p_j-1$.
\end{Thm}

%%%%%%%%%%%%%%%%%%%%%%%%%%%%%%%%%%%%%%%%%%%%%%%%%%%%%%%%%%%%%%%%%%%%%%
%%%                                                                %%%
%%%                         Summary                                %%%
%%%                                                                %%%
%%%%%%%%%%%%%%%%%%%%%%%%%%%%%%%%%%%%%%%%%%%%%%%%%%%%%%%%%%%%%%%%%%%%%%
\section{Summary}\label{Sec:summary}
We have considered the hidden subgroup problem
on the semi-direct product of cyclic groups $\Z_{N}\rtimes\Z_{p}$
with some restriction on $N$ and $p$.
By using the homomorphic properties,
we have presented a class of semi-direct product groups
in which the structures of subgroups can be easily classified.
Furthermore, we have shown that
there exists an efficient quantum algorithm
for the HSP on the class.

%%%%%%%%%%%%%%%%%%%%%%%%%%%%%%%%%%%%%%%%%%%%%%%%%%%%%%%%%%%%%%%%%%%%%%
%%%                                                                %%%
%%%                       Acknowledgements                         %%%
%%%                                                                %%%
%%%%%%%%%%%%%%%%%%%%%%%%%%%%%%%%%%%%%%%%%%%%%%%%%%%%%%%%%%%%%%%%%%%%%%
\acknowledgments{
D.P.C. was supported by a Korea Research Foundation grant (KRF-2004-059-C00060)
and by Asian Office of Aerospace Research and Development (AOARD-04-4003), and
S.L. was supported by the Kyung Hee University Research Fund in 2004 (KHU-20040915).
}

%%%%%%%%%%%%%%%%%%%%%%%%%%%%%%%%%%%%%%%%%%%%%%%%%%%%%%%%%%%%%%%%%%%%%%
%%%                                                                %%%
%%%                     Appendix                                   %%%
%%%                                                                %%%
%%%%%%%%%%%%%%%%%%%%%%%%%%%%%%%%%%%%%%%%%%%%%%%%%%%%%%%%%%%%%%%%%%%%%%
\appendix*
\section{}
In~\cite{inui}, Inui and Le Gall showed that there exists an
efficient quantum algorithm solving the HSP on the group
$\Z_{p^r}\rtimes\Z_p$.
In this appendix, we show that
Inui-Le Gall's method can be clearly applied to the case that
the group is $\Z_{2p^r}\rtimes\Z_p$, where $p$ is an odd prime.

We first note that
the multiplicative group $\Z_{2p^{r}}^{*}$ is cyclic and its order
is $p^{r-1}(p-1)$, and
that $\Z_{2p^r}\rtimes_\phi\Z_p$ is completely determined
by the elements of order $p$ in $\Z_{2p^{r}}^{*}$.
\begin{Prop}\label{Prop:order p elements}
 Let $p$ be an odd prime.
 Then $\alpha\in \Z_{2p^{r}}^{*}$ and $|\alpha|=p$
if and only if
\begin{equation}
\alpha=(2p^{r-1}+1)^{i}=2ip^{r-1}+1\pmod{2p^{r}},
\label{eq:Lem41}
\end{equation}
for $i\in \{1, 2,\ldots , p-1\}$.
\end{Prop}
\begin{proof}
It is clear that $(2p^{r-1}+1)^{ip}= 1 \pmod{2p^{r}}$ and
$(2ip^{r-1}+1)^{k}= 2ikp^{r-1}+1 \neq 1 \pmod{2p^{r}}$
for any $k \in \{1,2,\cdots ,p-1 \}$
since $p$ cannot divide $ik$.

Furthermore, since $\Z_{2p^{r}}^{*}$ is a cyclic group of order $p^{r-1}(p-1)$,
\begin{equation}
A=\{ 2ip^{r-1}+1 : i=0,1,2, \cdots, p-1 \}
\label{eq:subgroup_p}
\end{equation}
is a cyclic subgroup of $\Z_{2p^{r}}^{*}$.
Let $m$ be the minimal positive integer such that $ w^{m}\in A$
where $w$ is a generator of $\Z_{2p^{r}}^{*}$.
Suppose there exists $w^{b}\in\Z_{2p^{r}}^{*}$ such that $|w^{b}|=p$. % and  $w^{b} \notin A$.
Then since $b=sm+r$ for some $s, r \in \Z$ with $0\leq r<m$,
$(w^{r})^{p}=(w^{b})^{p}= 1 \pmod{2p^{r}}$.
Hence, by the minimality of $m$, we have $r=0$.
Therefore, $w^b\in A$, that is, every element of order $p$ in $\Z_{2p^{r}}^{*}$
is in $A$.
\end{proof}

%%%%%%%%%%%%%%%%%%%%%%%%%%%%%%%%%%%%%%%%%%%%%%%%%%%%%%%%%%%%%%%%%%%%%%%%%
%%%                                                                   %%%
%%%                Structure                                          %%%
%%%                                                                   %%%
%%%%%%%%%%%%%%%%%%%%%%%%%%%%%%%%%%%%%%%%%%%%%%%%%%%%%%%%%%%%%%%%%%%%%%%%%
\subsection{Structure of the semi-direct product group
 $\Z_{2p^{r}}\rtimes_{\phi} \Z_{p}$}\label{Sec:Structure}

Proposition~\ref{Prop:order p elements} implies that
$\phi(1)(1)\in A$ if and only if $\Z_{2p^{r}}\rtimes_{\phi} \Z_{p}$ is well-defined.
It follows that there are $p-1$ semi-direct product groups
of the form $\Z_{2p^{r}}\rtimes_{\phi} \Z_{p}$ other than $\Z_{2p^{r}}\times \Z_{p}$.

\begin{Prop}\label{Prop:Isomorphism}
For each $i=0,1,2,\ldots,p-1$,
let $\phi_i$ be a homomorphism from $\Z_{p}$ to $\mathrm{Aut}(\Z_{2p^{r}})$
defined as $\phi_i(1)(1)= (2p^{r-1}+1)^i$.
Then all $\Z_{2p^{r}}\rtimes_{\phi_i}\Z_{p}$'s are isomorphic to each other.
\end{Prop}
\begin{proof}
For each $i=1,2,\ldots,p-1$,
let $\Psi_i$ be a map
from $\Z_{2p^{r}}\rtimes_{\phi_1}\Z_p$ to $\Z_{2p^{r}}\rtimes_{\phi_i}\Z_p$
defined as $\Psi_i(x^ay^b)=x^ay^{bi^{-1}}$,
where $i^{-1}$ is the inverse of $i$ in $\Z_p^*$.
Then it can be easily checked that
$\Psi_i$ is a group isomorphism.
\end{proof}
By Proposition~\ref{Prop:Isomorphism},
it suffices to consider $\Z_{2p^{r}}\rtimes_{\phi}\Z_{p}$
with $\phi(1)(1)= 2p^{r-1}+1$.
Now, we formalize the group presentation
and the properties of the semi-direct product group $\Z_{2p^{r}}\rtimes_{\phi}\Z_{p}$,
and we then classify all of its subgroups % of $\Z_{2p^{r}}\rtimes_{\phi}\Z_{p}$
in terms of the group presentation.

The group presentation of %the semi-direct product group
$\Z_{2p^{r}}\rtimes_{\phi}\Z_{p}$ %with the homomorphism such that $\phi(1)(1)= 2p^{r-1}+1$
is
\begin{equation}
\left\langle x,y \mid
x^{2p^{r}}=y^{p}=e,~yx=x^{\phi(1)(1)}y=x^{2p^{r-1}+1}y\right\rangle.
\label{eq:gp_presentation}
\end{equation}
Then we can have

\begin{Prop}\label{Prop:equality}
\begin{eqnarray}
y^bx^a&=&x^{\phi(b)(a)}y^{b}=x^{a(2bp^{r-1}+1)}y^b,\nonumber \\
(x^ay^b)^k&=&x^{ak((k-1)bp^{r-1}+1)}y^{bk},\nonumber \\
(x^{a}y^{b})^{p}&=& x^{ap}
\label{eq:xyrelation}
\end{eqnarray}
and
\begin{equation}
|{x^ay^b}|=
\begin{cases}
\displaystyle\frac{2p^{r+1}}{\gcd(a,2p^r)}, &\textrm{ if $p^{r}|a$ and $b\neq 0$,}\\
\displaystyle\frac{2p^r}{\gcd(a,2p^r)}, &\textrm{ otherwise.}
\end{cases}
\label{eq:cases}
\end{equation}
for any $0\le a\le 2p^r-1$ and $0\le b\le p-1$.
\end{Prop}

\begin{proof}
It follows from straightforward calculations that the first three properties are true,
and for the last one, we divide it into several cases.

\noindent
{\bf Case 1. $p^{r}|a$ and $b \neq 0$.}
If $p^{r}|a$, then $a=0$ or $a=p^{r}$.

\noindent
(a) If $a=0$, then  $x^{a}y^{b}=y^{b}$.
Since $b \neq 0$,
\begin{equation}
|x^{a}y^{b}|=|y^{b}|=p=\frac{2p^{r+1}}{\gcd(a,2p^r)}.
\label{eq:prop20}
\end{equation}

\noindent
(b) If $a=p^{r}$, then  $x^{a}y^{b}=x^{p^{r}}y^{b}$.
Let $|x^{p^{r}}y^{b}|= d$. Then, by~(\ref{eq:xyrelation}), we have
\begin{equation}
(x^{p^{r}}y^{b})^{d}=x^{p^{r}d((d-1)bp^{r-1}+1)}y^{db} = e.
\label{eq:prop21}
\end{equation}
Since $b \neq 0$, $d=kp$ for some $k \in \Z$, and hence
\begin{equation}
(x^{p^{r}}y^{b})^{d}
=(x^{p^{r+1}})^{k}=e
\label{eq:prop22}
\end{equation}
which implies $2|k$.
Thus, $2p|d$ and $(x^{p^{r}}y^{b})^{2p}=e$, and hence
\begin{equation}
d=2p=\frac{2p^{r+1}}{\gcd(a,2p^r)}.
\label{eq:prop23}
\end{equation}

\noindent
{\bf Case 2. $p^{r}\nmid a$ or $b=0$.}

\noindent
(a) If $b=0$ then it is clear that
\begin{equation}
\left|x^a y^b\right|=\left|x^a\right|=\frac{2p^{r}}{\gcd(a,2p^r)}.
\label{eq:prop24}
\end{equation}

\noindent
(b) If $b\neq 0$ then $p^{r}\nmid a$.
Let $|x^{a}y^{b}|=d$.
Then since
\begin{equation}
(x^{a}y^{b})^{d}=x^{a d((d-1)bp^{r-1}+1)}y^{db} = e,
\label{eq:prop25}
\end{equation}
we have $p|d$ and so $(x^ay^b)^d=x^{ad}=(x^{a})^d=e$.
Thus, we have
\begin{equation}
\left.\frac{2p^{r}}{\gcd(a,2p^r)}\right| d.
\label{eq:prop26}
\end{equation}
Since $p^{r}\nmid a$, the left-hand side of (\ref{eq:prop26}) is a multiple of $p$,
and hence
\begin{equation}
(x^{a}y^{b})^{{2p^{r}}/{\gcd(a,2p^r)}}=x^{a{2p^{r}}/{\gcd(a,2p^r)}}=e.
\label{eq:prop260}
\end{equation}
Therefore, by the minimality of $d$, we obtain
\begin{equation}
d=\frac{2p^{r}}{\gcd(a,2p^r)}.
\label{eq:prop261}
\end{equation}

\end{proof}

Furthermore, for cyclic subgroups generated by the element
$x^ay^b$, we can have the following property.

\begin{Prop}\label{Prop:aibi}
Let $0\le a\le 2p^r-1$ and $0\le b\le p-1$.
If $p^{r}|a$ and $b \neq 0$ then
\begin{equation}
\left\langle{x^ay^b}\right\rangle=
\left\{x^{ai}y^{bi} : 0\le i\le \frac{2p^{r+1}}{\gcd(a,2p^r)}-1\right\}.
\label{eq:prop30}
\end{equation}
Otherwise,
\begin{equation}
\left\langle{x^ay^b}\right\rangle=
\left\{x^{ai}y^{bi} : 0\le i\le \frac{2p^{r}}{\gcd(a,2p^r)}-1\right\}.
\label{eq:prop31}
\end{equation}
\end{Prop}

\begin{proof}
Let $i$ be an arbitrary integer. Then, by~(\ref{eq:xyrelation}),
\begin{eqnarray}
(x^{a}y^{b})^{i}&=&x^{ai(i-1)bp^{r-1}}x^{ai}y^{bi}\nonumber\\
&=&((x^{a}y^{b})^{p})^{i(i-1)bp^{r-2}}x^{ai}y^{bi}.
\label{eq:prop32}
\end{eqnarray}
Thus
\begin{equation}
\left(((x^{a}y^{b})^{p})^{i(i-1)bp^{r-2}}\right)^{-1}(x^{a}y^{b})^{i}=x^{ai}y^{bi},
\label{eq:prop33}
\end{equation}
and hence
\begin{equation}
\left\{x^{ai}y^{bi} : i\in \Z\right\}\subseteq \left\langle x^{a}y^{b}\right\rangle.
\label{eq:prop34}
\end{equation}
Furthermore, $x^{ai}y^{bi}\neq x^{aj}y^{bj},$ for any $i, j$ with
$1\le i< j\le \frac{2p^{r+1}}{\gcd(a,2p^r)}-1$ if $p^{r}|a$ and $b \neq 0$,
and with $1\le i<j\le \frac{2p^{r}}{\gcd(a,2p^r)}-1$ otherwise.
Hence, we obtain (\ref{eq:prop30}) and (\ref{eq:prop31}).
\end{proof}

%%%%%%%%%%%%%%%%%%%%%%%%%%%%%%%%%%%%%%%%%%%%%%%%%%%%%%%%%%%%%%%%%%%%%%%%%
%%%                                                                   %%%
%%%                classification of subgroup                         %%%
%%%                                                                   %%%
%%%%%%%%%%%%%%%%%%%%%%%%%%%%%%%%%%%%%%%%%%%%%%%%%%%%%%%%%%%%%%%%%%%%%%%%%

%\subsection{Classification of subgroups $H \subseteq~
%\Z_{2p^{r}}\rtimes_{\phi}\Z_{p}$}

Now, we are ready to classify all possible subgroups of
$\Z_{2p^{r}}\rtimes_{\phi}\Z_{p}$ in terms of group presentation.
Let $H'=H \cap \left\langle x \right\rangle$.
Then $H'=\left\langle x^{2^{t}p^{s}}\right\rangle$
for some integers $t$ and $s$ with $0\le t\le 1$ and $0\le s \le r$.

We assume that $H\neq\left\langle x^{2^{t}p^{s}}\right\rangle$.
Then it is clear that
$x^{a_0}y^{b_0}\in H$ for some integers $a_0$ and $b_0$
with $0\le a_0\le 2p^r-1$ and $1\le b_0\le p-1$.
By Proposition~\ref{Prop:aibi},
$x^{a_0 b_0^{-1}} y \in \left\langle x^{a_0}y^{b_0}\right\rangle \subseteq H$,
where $b_0^{-1}$ is the multiplicative inverse of $b_0$ in $\Z_{p}^{*}$.
Furthermore, since $(x^{a_0b_0^{-1}} y)^p= x^{a_0b_0^{-1}p}\in H$,
$2^{t}p^{s-1} | a_0b_0^{-1}$.
Thus, we have $x^{h2^{t}p^{s-1}}y\in H$
where $h$ is an integer
such that $h2^{t}p^{s-1}=a_0b_0^{-1} \pmod{2^tp^s}$ and $0\le h\le p-1$.
Hence we obtain
$\left\langle x^{2^{t}p^{s}} , x^{h2^{t}p^{s-1}}y \right\rangle \subseteq H$.

Let $x^{a}y^{b} \in H$ with $0\le a\le 2p^r-1$ and $0\le b\le p-1$.
Then since $x^{h2^{t}p^{s-1}b}y^{b}\in \left\langle x^{h2^{t}p^{s-1}}y\right\rangle$
by Proposition~\ref{Prop:aibi},
its inverse $(x^{h2^{t}p^{s-1}b}y^{b})^{-1}=y^{-b}x^{-h2^{t}p^{s-1}b}$ is also in
$\left\langle x^{h2^{t}p^{s-1}}y\right\rangle\subseteq H$.
Thus, $(x^a y^{b})(y^{-b}x^{-h2^{t}p^{s-1}b})=x^{a-h2^{t}p^{s-1}b}$ is contained in $H$,
and so it is in $H'$.
Hence we obtain $2^tp^s | a-h2^{t}p^{s-1}b$, that is,
$a=2^tp^s m +h2^{t}p^{s-1}b$ for some integer $m$.
Thus,
\begin{equation}
x^ay^b=\left(x^{2^tp^s}\right)^m \left(x^{h2^{t}p^{s-1}b}y^b\right)\in
\left\langle x^{2^{t}p^{s}} , x^{h2^{t}p^{s-1}}y \right\rangle,
\label{eq:classify11}
\end{equation}
that is, $H=\left\langle x^{2^{t}p^{s}} , x^{h2^{t}p^{s-1}}y \right\rangle$.

Therefore we obtain the following proposition.

\begin{Prop}\label{Prop:subgroups}
 For any subgroup $H\in \Z_{2p^{r}}\rtimes_{\phi}\Z_{p}$,
 $H$ is one of $\left\langle x^{2^{t}p^{s}}\right\rangle$,
 or $\left\langle x^{2^{t}p^{s}},~x^{h2^{t}p^{s-1}}y\right\rangle$
 for some integers $t$, $s$, and $h$ with $0\le t\le 1$, $0\le s \le r$, and $0\le h \le p-1$.
\end{Prop}

\begin{Prop}\label{Prop:subset}
Let $t$, $s$, and $h$ be integers with $0\le t\le 1$, $0\le s \le r$, and $0\le h \le p-1$.
Then
\begin{displaymath}
\left\{x^{(bh \bmod p)2^{t}p^{s-1}}y^{b}\right\}_{0\leq b \leq p-1}
\subseteq \left\langle x^{2^{t}p^{s}},~x^{h2^{t}p^{s-1}}y\right\rangle.
\end{displaymath}
\end{Prop}

\begin{proof}
For any $b=0, \cdots, p-1$, and $h=0, \cdots, p-1$,
\begin{equation}
x^{bh2^{t}p^{s-1}}y^{b}=x^{l2^{t}p^{s}}x^{(bh \bmod p)2^{t}p^{s-1}}y^{b},
\label{eq:Prop_subset}
\end{equation}
where $l$ is an integer such that $bh=lp+(bh \bmod p)$.
Since $x^{l2^{t}p^{s}}\in \left\langle x^{2^{t}p^{s}}\right\rangle$
and $x^{bh2^{t}p^{s-1}}y^{b}\in\left\langle x^{h2^{t}p^{s-1}}y\right\rangle $
by Proposition~\ref{Prop:aibi},
\begin{equation}
x^{(bh \bmod p)2^{t}p^{s-1}}y^{b}=(x^{l2^{t}p^{s}})^{-1}x^{bh2^{t}p^{s-1}}y^{b}
\label{eq:Prop_subset2}
\end{equation}
is in $\left\langle x^{2^{t}p^{s}}, x^{h2^{t}p^{s-1}}y\right\rangle$.

\end{proof}

%%%%%%%%%%%%%%%%%%%%%%%%%%%%%%%%%%%%%%%%%%%%%%%%%%%%%%%%%%%%%%%%%%%%%%%%%
%%%                                                                   %%%
%%%                Quantum algorithm                                  %%%
%%%                                                                   %%%
%%%%%%%%%%%%%%%%%%%%%%%%%%%%%%%%%%%%%%%%%%%%%%%%%%%%%%%%%%%%%%%%%%%%%%%%%
\subsection{Quantum algorithm}\label{Sec:Algo}
Here, we present a quantum algorithm solving the HSP over $\Z_{2p^{r}}\rtimes_{\phi}\Z_{p}$,
which is the same as that in~\cite{inui}.
The hidden subgroup of $\Z_{2p^{r}}\rtimes_{\phi}\Z_{p}$
with respect to the oracle function $f$ is denoted by $H$.
Then the procedure is as follows.
First, employing the abelian HSP algorithm on a cyclic group $\left\langle x \right\rangle$,
we find $H'=H\cap \left\langle x \right\rangle=\left\langle x^{2^{t}p^{s}}\right\rangle$,
and then
we determine whether $H=\left\langle x^{2^{t}p^{s}}\right\rangle$
or $H=\left\langle x^{2^{t}p^{s}}, x^{h2^{t}p^{s-1}}y\right\rangle$,
and find $h$ if $H=\left\langle x^{2^{t}p^{s}}, x^{h2^{t}p^{s-1}}y\right\rangle$,
by means of the following quantum algorithm:

%Quantum~~ Algorithm
\noindent
1. Prepare the state
\begin{equation}
\frac 1 p \sum_{a=0}^{p-1}\sum_{b=0}^{p-1}\ket{a}\ket{b}\ket{f(x^{a2^{t}p^{s-1}}y^b)}.
\label{eq:algorithm}
\end{equation}
2. Measure the third register.

\noindent
3. Apply $\cF_p \otimes \cF_p$ to the first two
registers, where $\cF_p$ is the quantum Fourier transform over
$\Z_p$, that is, $\iota=\sqrt{-1}$ and
\begin{equation}
\cF_p |l\rangle=\frac 1 {\sqrt p}\sum_{k=0}^{p-1} e^{2\pi \iota {kl}/ p}
\ket{k}~\textrm{ for all }~0\le l\le p-1.
\label{eq:Fourier}
\end{equation}

\noindent
4. Measure the first and the second register:
we get two values $\tilde c$ and $\tilde d$.
If $\tilde c=0$, then we regard the procedure as failed.
Otherwise, compute $\tilde h=-\tilde c^{p-2}\tilde d \pmod p$ and output $\tilde h$.

We repeat this procedure $k=\Theta(1)$ times,
where the details are in the next section.
If we obtain the same $\tilde h$ at every repetition,
we conclude that $H = \left\langle x^{2^{t}p^{s}}, x^{h2^{t}p^{s-1}}y \right\rangle$ with $h=\tilde h$.
If we obtain at least two different values for $\tilde h$ during the $k$ repetitions,
we conclude that $H=H'$.
Furthermore, we can readily show that
the total time complexity is $O((r\log p)^2)$.

%%%%%%%%%%%%%%%%%%%%%%%%%%%%%%%%%%%%%%%%%%%%%%%%%%%%%%%%%%%%%%%%%%%%%%%%%
%%%                                                                   %%%
%%%                Analysis of the algorithm                          %%%
%%%                                                                   %%%
%%%%%%%%%%%%%%%%%%%%%%%%%%%%%%%%%%%%%%%%%%%%%%%%%%%%%%%%%%%%%%%%%%%%%%%%%
\subsection{Analysis of the algorithm}\label{Sec:Analysis}
In this section, we show that we can find $H$ with sufficiently high
probability by the above algorithm.

%%%                                                                   %%%
%%% The case when H=\langle{x^{2^{t}p^{s}},~x^{h2^{t}p^{s-1}}y}\rangle%%%
%%%                                                                   %%%
\subsubsection{The case when $H=\left\langle{x^{2^{t}p^{s}},~x^{h2^{t}p^{s-1}}y}\right\rangle$}

For the initial state
\begin{equation}
\frac 1 p \sum_{a=0}^{p-1}\sum_{b=0}^{p-1}\ket{a}\ket{b}
\ket{f(x^{a2^{t}p^{s-1}}y^b)},
\label{eq:analysis01}
\end{equation}
take $a_0=a-bh \bmod p\in \Z_p$.
Then the state in (\ref{eq:analysis01}) can be rewritten by use of
the summation on $a_{0}$ instead of the summation on $a$ as follows.

\begin{align}
\frac 1 p \sum_{a_{0}=0}^{p-1}\sum_{b=0}^{p-1}&
\ket{a_{0}+ bh \bmod p} \ket{b}\nonumber\\
& \ket{f(x^{a_{0}2^{t}p^{s-1}}x^{(bh \bmod p){2^{t}p^{s-1}}}y^b)}.
\label{eq:analysis02}
\end{align}

By Proposition~\ref{Prop:subset}, for each integer $b$ with $0\le b\le p-1$,
the element $x^{(bh \bmod p)2^{t}p^{s-1}}y^b$ is in the hidden subgroup $H$,
and thus all the elements
$x^{a_{0}2^{t}p^{s-1}}x^{(bh \bmod p)2^{t}p^{s-1}}y^b$
are mapped to the same value, that is,
\begin{equation}
f(x^{a_{0}2^{t}p^{s-1}}x^{(bh \bmod p)2^{t}p^{s-1}}y^b)=f(x^{a_{0}2^{t}p^{s-1}}),
\label{eq:analysis05}
\end{equation}
for any $0\le a_0, b\le p-1$.
Thus, discarding the third register after the second step,
the resulting state becomes
\begin{equation}
\ket{\psi}=\frac {1}{\sqrt p} \sum_{b=0}^{p-1}\ket{a_0+bh \bmod p}\ket{b},
\label{eq:analysis06}
\end{equation}
where $a_0$ is randomly determined by the measurement.

By applying the quantum Fourier transform to $\ket{\psi}$,
the state becomes
\begin{eqnarray}
\cF_p \otimes \cF_p \ket{\psi}
&=&\frac 1{p\sqrt{p}}\sum_{c,d=0}^{p-1} e^{2\pi \iota {a_{0}c}/{p}}
\sum_{b=0}^{p-1}e^{2\pi \iota b({ch+d})/{p}}\ket{c}\ket{d}\nonumber \\
&=&\frac{1}{\sqrt{p}}\sum_{p|ch+d}%{\begin{subarray}{c}c,d\\ p|ch+d  \end{subarray}}
e^{2\pi \iota {a_{0}c}/{p}}\ket{c}\ket{d}.
\label{eq:analysis07}
\end{eqnarray}
Since, after the fourth step,
the measured values $\tilde c$ and $\tilde d$ satisfy $p|\tilde ch+\tilde d$,
if $\tilde c\ne0$ then we obtain $\tilde h$
by computing $-\tilde c^{p-2}\tilde d \bmod p$.
Hence, with probability $1-1/p$, we obtain the value of $h$.

%%%                                                                   %%%
%%%          The case when $H=\langle x^{2^{t}p^{s}}\rangle$          %%%
%%%                                                                   %%%
\subsubsection{The case when $H=\left\langle x^{2^{t}p^{s}}\right\rangle$}
In this case, we note that %for any $b, b' \in \{0,1, \cdots, p-1 \}$,
if $x^{a}y^{b}H=x^{a}y^{b'}H$ then $b=b'$,
and that if $x^{a2^{t}p^{s-1}}y^{b}H=x^{a'2^{t}p^{s-1}}y^{b}H$ then
since $x^{a2^{t}p^{s-1}}y^{b}(x^{a'2^{t}p^{s-1}}y^{b})^{-1}=x^{(a-a')2^{t}p^{s-1}}\in H$,
we have $p\mid a-a'$, which implies $a=a'$.
Thus, the oracle function $f$ is an injective map on the set
$\left\{x^{a2^{t}p^{s-1}}y^b : 0 \leq a,b \leq p-1 \right\}$.

After measuring and discarding the third register, the state is
$\ket{\psi'}=\ket{a_0}\ket{b_0}$,
where $a_{0}$ and $b_{0}$ are determined
by the measurement outcome of the third register $\ket{f(x^{a_{0}2^{t}p^{s-1}}y^{b_0})}$.

By applying quantum Fourier transform, we can have
\begin{equation}
\cF_p \otimes \cF_p \ket{\psi'}
=\frac{1}{p}\sum_{c,d=0}^{p-1} e^{2\pi \iota (a_{0}c+b_{0}d)/p}
\ket{c}\ket{d},
\label{eq:analysis08}
\end{equation}
which is a uniform superposition of the values $c$ and $d$ in $\Z_p$.

If the measurement outcome of the first register is zero, that is, $\tilde c= 0$,
then we disregard the result.
If $\tilde c$ is not zero, then
$\{\tilde h=-\tilde c^{p-2}\tilde d \bmod p\}$ forms a uniform distribution over $\Z_p$.
Therefore, with high probability, we obtain at least two different values $\tilde h$'s.

%%%%%%%%%%%%%%%%%%%%%%%%%%%%%%%%%%%%%%%%%%%%%%%%%%%%%%%%%%%%%%%%%%%%%%%%%
%%%                                                                   %%%
%%%                Success probability                                %%%
%%%                                                                   %%%
%%%%%%%%%%%%%%%%%%%%%%%%%%%%%%%%%%%%%%%%%%%%%%%%%%%%%%%%%%%%%%%%%%%%%%%%%
\subsubsection{Success probability}\label{Sec:prob}
We now consider the success probability
when repeating the procedure in the previous section $k$ times,
after obtaining $t$ and $s$ by the abelian HSP algorithm.

If $H=\left\langle x^{2^{t}p^{s}}, x^{h2^{t}p^{s-1}}y\right\rangle$,
then, with probability $1/p^k$, it fails to output the correct $h$.

If $H=\left\langle x^{2^{t}p^{s}}\right\rangle$,
then the probability that $\tilde{c}=0$ in the fourth step for every repetition
is $1/p^k$.
In addition, although all values of $\tilde{c}$'s are not zero,
the probability of incorrect output,
that is, the probability of deciding that
$H$ is $H=\left\langle x^{2^{t}p^{s}},~x^{h2^{t}p^{s-1}}y\right\rangle$
for a value $0\le h \le p-1$,
is $(2^k-1)/p^{k-1}$, which can be easily shown.

Therefore, the total success probability is at least
$1-(2^kp-p+1)/p^{k}$, and hence
taking some proper constant $k$,
we can obtain the correct $h$ with high probability by $k$ repetitions.


\begin{thebibliography}{1}

\bibitem{Simon}
D.~Simon,
% {\em On the Power of Quantum Computation},
 Proceedings of the 35th Annual IEEE Symposium
 on the Foundations of Computer Science (Piscataway, NJ),
 IEEE Computer Society Press,  pp. 116--123, 1994;
 SIAM Journal on Computing, {\bf 26}, pp. 1474--1483 (1997).

\bibitem{Shor}
 P. W. Shor,
% {\em Algorithms for Quantum Computation:
% Discrete Logarithms and Factoring},
 Proceedings of the 35th Annual IEEE Symposium
 on the Foundations of Computer Science (Piscataway, NJ),
 IEEE Computer Society Press, pp. 124--134, 1994;
 SIAM Journal on Computing, {\bf 26}, pp. 1484--1509 (1997).

\bibitem{iso}
M.~Ettinger and P.~H{\o}yer,
%{\em A quantum observable for the graph isomorphism problem},
quant-ph/9901029, 1999.

\bibitem{lattice}
O.~Regev,
%{\em Quantum computation and lattice problems},
 Proceedings of the 43rd Annual IEEE Symposium
 on the Foundations of Computer Science,
 IEEE Computer Society Press, pp.520--529 (2002);
 arXiv:cs.DS/0304005, 2003.

\bibitem{RB}
M.~R\"{o}tteler and T.~Beth,
%{\em Polynomial-time solution to the hidden subgroup problem
%for a class of non-abelian groups},
quant-ph/9812070, 1998.

\bibitem{ettinger}
M.~Ettinger and P.~H{\o}yer,
%{\em On Quantum Algorithms for Non Commutative Hidden Subgroups},
Advances in Applied Mathematics, {\bf 25} pp. 239--251 (2000).

\bibitem{HRT}
S.~Hallgren, A.~Russell, and A.~Ta-Shma,
%{\em Normal subgroup reconstruction and quantum computing using group representations},
Proceedings of the 32nd Annual ACM Symposium on Theory of Computing, pp. 627--635 (2000).

\bibitem{GSVV}
M.~Grigni, L.~Schulman, M.~Vazirani, and U.~Vazirani,
%{\em Quantum mechanical algorithms for the nonabelian hidden subgroup problem},
Proceedings of the 33rd Annual ACM Symposium on Theory of Computing, pp. 68--74 (2001).

\bibitem{friedl}
K.~Friedl, G.~Ivanyos, F.~Magniez, M.~Santha, and P.~Sen,
%{\em Hidden Translation and Orbit Coset in Quantum Computing},
Proceedings of the 35th Annual ACM Symposium on Theory of Computing,
pp. 1--9 (2003).

\bibitem{kuperberg}
G.~Kuperberg,
%{\em A Subexponential-time Quantum Algorithm for the Dihedral Subgroup Problem},
quant-ph/0302112, 2003.

\bibitem{EHN}
M.~Ettinger, P.~H{\o}yer, and E.~Knill,
%{\em The quantum query complexity of the hidden subgroup problem is polynomial},
Inform. Process. Lett., {\bf 91} pp. 43--48 (2004);
quant-ph/0401083, 2004.

\bibitem{gavinsky}
D.~Gavinsky,
%{\em Quantum solution to the hidden subgroup problem for poly-near-hamiltonian groups},
Quantum Inf. Comput., {\bf 4} pp. 229--235 (2004).

\bibitem{inui}
Y.~Inui and F.~Le~Gall,
%{\em An efficient algorithm for the hidden subgroup problem
%over a class of semi-direct product groups},
quant-ph/0412033, 2004.

\bibitem{moore}
C.~Moore, D.~N. Rockmore, A.~Russell and L.~J. Schulman,
%{\em The Power of Basis Selection in Fourier Sampling: Hidden Subgroup Problems in Affine Groups},
Proceedings of the 15th Annual ACM-SIAM Symposium on Discrete Algorithms, pp. 1106--1115 (2004).

\bibitem{regev}
O.~Regev,
%{\em A Subexponential Time Algorithm for the Dihedral Hidden Subgroup Problem with Polynomial Space},
quant-ph/0406151, 2004.

\bibitem{BCD}
D.~Bacon, A.~M.~Childs, and W.~van~Dam,
%{\em From optimal measurement to efficient quantum algorithms
%for the hidden subgroup problem over semidirect product groups},
Proc. 46th IEEE Symposium on Foundations of Computer Science (FOCS 2005), pp. 469--478 (2005);
quant-ph/0504083, 2005.

\bibitem{BCD2}
D.~Bacon, A.~M.~Childs, and W.~van~Dam,
%{\em Optimal measurements for the dihedral hidden subgroup problem},
quant-ph/0501044, 2005.

\end{thebibliography}
\end{document}